# A PMU Based Islanding Detection Scheme Immune to Additive Instrumentation Channel Errors

Meghna Barkakati, *Member, IEEE,* Reetam Sen Biswas, *Student Member*, *IEEE* and Anamitra Pal, *Member, IEEE*
School of Electrical, Computer, and Energy Engineering
Arizona State University, Tempe, Arizona-85287, USA
Email: mbarkaka@asu.edu, rsenbisw@asu.edu, anamitra.pal@asu.edu

*Abstract*— Traditional synchrophasor measurement-based islanding detection techniques have primarily relied on voltage angle measurements and/or their derivatives for successfully detecting islands. However, relatively high instrumentation channel errors associated with phasor measurement unit (PMU) data, can significantly degrade islanding detection accuracies. In this paper, a new islanding detection scheme employing cumulative sum of change in voltage phase angle difference (CUSPAD) is proposed, which is immune to additive instrumentation channel errors in the PMU measurements. The robustness of the proposed islanding detection algorithm is established through application to an 18-bus test system and the IEEE 118-bus system having different wind energy penetration levels. Comparative analysis of the accuracies of the proposed approach (CUSPAD) and the conventional angle difference (AD) approach prove the former's superior performance when additive instrumentation channel errors are present.

## I. INTRODUCTION

Unintentional power system islanding refers to an uncontrolled separation of a portion of the electrical network from the rest of the system. It can occur due to power system disturbances (such as faults), natural events (such as hurricanes), or human mis-operation [1]-[3]. Fast and accurate detection of an island when it has formed is essential for the prompt restoration of the system. The role of phasor measurement units (PMUs) in detecting, identifying, maintaining, and eventually restoring the system after the 2008 Hurricane Gustav has been described in [2], [3]. Frequency measurements from PMUs obtained during Hurricane Gustav helped operators monitor the island's load generation balance by adjusting governor controls, which prevented system collapse.

PMUs provide time-synchronized information of complex voltage and current phasors, frequency, and rate-of-change-of frequency. References [4] and [5] have used frequency differences and voltage phase angle differences for islanding detection, respectively. Principal component analysis (PCA) on voltage magnitudes, phase angles, and frequency measurements have been investigated for reliable islanding detection in [5], [6]. Data mining techniques such as support vector machine (SVM) and decision trees (DTs) were applied for islanding detection in [7] and [8], respectively. In [9], a comparative study revealed that DT based classifiers were most dependable for passive islanding detection. Additionally, Tokyo Electric Power Company, Inc. (TEPCO) performed a comparative analysis of different attributes for islanding detection and acknowledged that phase angle difference obtained from PMUs is the most reliable method for detecting unintentional islanding [10].

PMU measurements are however susceptible to both device errors as well as instrumentation channel errors. As per [11], PMU device error expressed as a total vector error (TVE) is typically less than 1%. However, the errors introduced by the instrumentation channel may cause a phase-shift that can be as high as $\pm 4°$ [12]. Thus, reliable and fast detection of unintentional islanding in the presence of instrumentation errors in PMU measurements can be a major challenge [13]. Considering the recent advancements made in renewable energy generation technology, the contribution of inverter-based generation (IBG) such as wind and solar in the transmission network is expected to increase significantly in the near-future. Additionally, during transients a high renewable energy penetration may have a substantial impact on system stability [14]. Prior research on islanding detection considering renewable energy penetration has primarily focused on distribution grids [15], [16]. Islanding detection in IBG-dominated transmission grids is important because when a renewable-rich sub-system gets isolated from the bulk power system, power quality issues such as frequency deviation, voltage fluctuation, and harmonics may manifest as critical problems. Therefore, it is important to detect unintentional islanding quickly and take immediate corrective actions (such as fast tripping of the isolated IBG). The research problem being explored in this paper can be framed as: *accurately detect unintentional islanding using PMUs in the transmission grid in presence of IBG and additive instrumentation channel errors.*

The rest of the paper is organized as follows. Section II describes the proposed islanding detection scheme. Section III presents the simulation results. Section IV summarizes the major contributions and concludes the paper.

## II. PROPOSED ISLANDING DETECTION METHODLOGY

This section introduces the need for a new PMU-based islanding detection scheme. The reason why the proposed

This work was supported by the PSERC project S-74

technique is immune to additive instrumentation channel errors is explained next. The methodology followed for modeling wind energy penetration using a positive sequence simulation software is described afterwards. Finally, this section concludes by describing a supervised learning scheme using DTs as well as the methodology that was employed for placing the PMUs.

*A. Need for a new islanding detection scheme*

Let the true bus voltage angles at any two buses $i$ and $j$ at time instant t be given by $\theta_i^t$ and $\theta_j^t$, respectively. The traditional angle difference (AD) approach for islanding detection computes the difference between $\theta_i^t$ and $\theta_j^t$ [17] as shown below:

$$\Delta \theta^t = \theta_i^t - \theta_j^t \qquad (1)$$

When the calculated voltage angle difference, $\Delta \theta^t$ exceeds a pre-determined threshold, $\tau$, the approach concludes that an island has formed. It is worth mentioning here that $\tau$ is often obtained from offline analyses that do not account for the actual errors present in the system. Let the instrumentation channel errors associated with the phase angle measurements at buses $i$ and $j$ be $e_i$ and $e_j$, respectively. Then, the measured angle differences at buses $i$ and $j$ are given by $\theta_i^m = \theta_i^t + e_i$ and $\theta_j^m = \theta_j^t + e_j$, respectively. Consequently, the measured angle difference between buses $i$ and $j$, is:

$$\Delta \theta^m = \theta_i^m - \theta_j^m = (\theta_i^t + e_i) - (\theta_j^t + e_j) = \Delta \theta^t + (e_i - e_j) \qquad (2)$$

Due to the error $(e_i - e_j)$ in the measured voltage angle $\Delta \theta^m$, the following situations may occur:

*1)* $\Delta \theta^t > \tau$, but $\Delta \theta^m < \tau$: In this scenario, an unintentional islanding may not be detected.

*2)* $\Delta \theta^t < \tau$, but $\Delta \theta^m > \tau$: In this scenario, a non-islanding contingency may be misclassified as unintentional islanding.

In light of the two scenarios mentioned above, it is clear that the accuracy of the conventional AD approach would decrease in presence of large instrumentation channel error.

*B. Input feature for islanding detection*

As described in Section II.A, conventional AD approach for islanding detection may not be reliable for detecting unintentional islanding in presence of instrumentation channel errors. The major contribution of this paper is the development of a pre-processing technique on the input feature set that makes the detection methodology immune to fixed additive instrumentation errors. In our case, the input features are the voltage phase angles obtained from PMUs. We do this by first stating (and proving) the following lemma.

**Lemma 1:** *Cumulative sum of change in voltage phase angle computed with respect to a pre-contingency reference angle obtained from the same PMU device over a given time-period is immune to instrumentation channel errors.*

*Proof:* Let $\theta_x^t$ and $\theta_x^m$ denote the true voltage angle and the measured voltage angle, respectively, i.e. $\theta_x^m = \theta_x^t + e_x$ holds true for every time instant, where $e_x$ denotes the fixed but unknown instrumentation error [18]. Now, let a contingency occur at $t = t_c$ that causes the voltage angles to change in the manner shown in Fig. 1. Note that the pre-contingency voltage angle is the reference voltage angle, denoted by $\theta_x^t(t_c^-)$ for the true angle and $\theta_x^m(t_c^-)$ for the measured angle, respectively, where $\theta_x^m(t_c^-) = \theta_x^t(t_c^-) + e_x$. The cumulative sum of change in voltage phase angles for the true angle and the measured angle are denoted by the green and the blue shaded regions and can be mathematically written as:

$$\left.\begin{aligned} S_x^t &= \sum_{n=1}^{w} |\theta_x^t(t_c + n) - \theta_x^t(t_c^-)| \\ S_x^m &= \sum_{n=1}^{w} |\theta_x^m(t_c + n) - \theta_x^m(t_c^-)| \end{aligned}\right\} \qquad (3)$$

Now, as the additive instrumentation error is an unknown but fixed quantity, they will cancel out at every time instant of $S_x^m$ making it equal to $S_x^t$. Therefore, although the true and the measured voltage phase angles are numerically different (the blue and green curves have different intercepts on the Y-axis), the area under the curve between the post-contingency voltage angle and the reference voltage angle (over a window of w samples) for both true and measured voltage angles will be the same. In other words, the following holds true:

$$\left.\begin{aligned} \theta_x^m &\neq \theta_x^t \\ S_x^m &= S_x^t \end{aligned}\right\} \qquad (4)$$

From (4) it can be concluded that the cumulated sum of change in voltage phase angle obtained from a specific PMU over a time trajectory does not get affected by instrumentation channel errors. This proves Lemma 1.

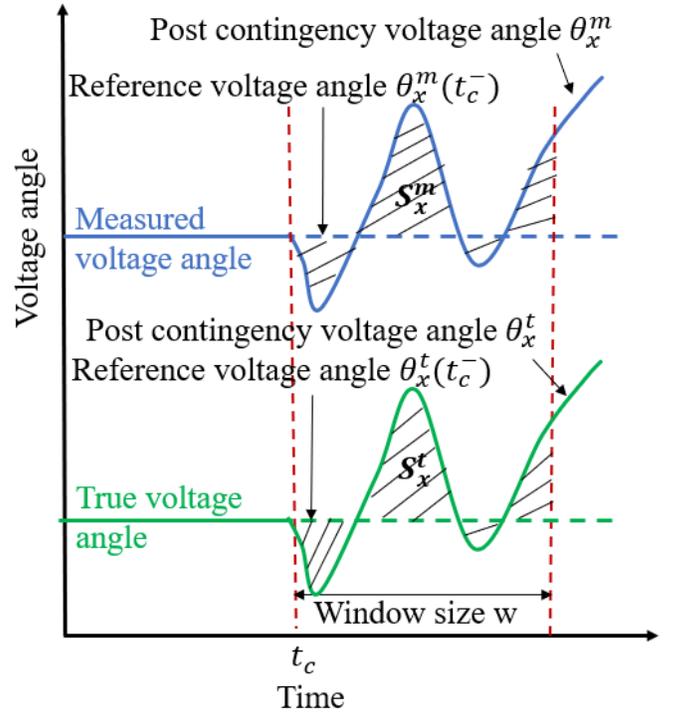

Fig. 1: Schematic diagram depicting immunity of CUSPAD to additive instrumentation errors.

The traditional approach for detecting islands used the raw angle differences between two buses, say, $x$ and $y$, given by

$\theta_x^m - \theta_y^m$, as input feature for decision-making. Considering Lemma 1, in this paper, the following methodology is devised for selecting the input feature. If a contingency occurs at $t = t_c$, the cumulative sum of change in voltage phase angles for buses $x$ and $y$, denoted by $S_x^m$ and $S_y^m$ is computed based on the relation shown in (3). Since, $S_x^m$ and $S_y^m$ are immune to instrumentation error, the input feature for islanding detection is chosen to be the cumulative sum of change in voltage phase angle difference (CUSPAD) between buses *x* and *y*, which is mathematically described by:

$$CUSPAD_{xy} = S_x^m - S_y^m \quad (5)$$

However, for real-time applications the *determination of the pre-contingency reference voltage angle $\theta_x(t_c^-)$ and $\theta_y(t_c^-)$ in real-time* is a concern. This problem can be resolved by using the three-sample based quadratic prediction algorithm (TSQPA) proposed by Gao et al. in [19], and extended to multiple load models in [20]. TSQPA states that for a linear change in load (which is a valid assumption to make considering the fast output rates of PMUs), the relationship between successive voltages is given by:

$$V(n|n-1) = 3V(n-1) - 3V(n-2) + V(n-3) \quad (6)$$

where, $V(n|n-1)$ denotes the predicted value of complex voltage at time instant n, when the voltages at time instants $n-3$ through $n-1$ are known. From the predicted value of the complex voltage, $V(n|n-1)$, the predicted voltage phase angle, $\theta(n|n-1)$, can be obtained. Knowing the predicted phase angle, $\theta(n|n-1)$, and the measured phase angle, $\theta(n)$, an observation residual, $r(n)$, can be computed as follows:

$$r(n) = \theta(n|n-1) - \theta(n) \quad (7)$$

When the observation residual, $r(n)$, manifests a sudden change, it means a contingency has occurred at time instant $n$ and the reference voltage angle for CUSPAD calculation must be the angle just before that time instant, i.e., $\theta(n-1)$. Based on the analysis done above, the main result of this paper is described by the following theorem.

**Theorem 1:** *For islanding detection in presence of additive instrumentation errors, a CUSPAD-based approach has higher accuracy than the conventional angle difference (AD)-based approach.*

*Proof*: Section II.A demonstrates how islanding detection accuracy of the conventional AD approach would deteriorate in presence of additive instrumentation channel errors. Lemma 1 proves how CUSPAD computed with respect to a pre-contingency reference angle becomes immune to additive instrumentation channel errors. By combining the two arguments it can be concluded that CUSPAD will provide better performance in comparison to the conventional AD method for islanding detection in presence of large instrumentation channel errors. This proves Theorem 1.

### C. Wind energy modeling

A wind farm is a collective group of interconnected wind turbines that are tied to a point of common coupling (PCC) before the power is fed to the grid. In accordance with the WECC Wind Plant Power Flow Modeling Guide, wind power plants must be represented by an equivalent generator, generator transformer, collector system, and substation transformer [21]. The characteristic features of the wind farm used in this study are described below.

A wind farm containing several wind turbines is modeled as an equivalent generator as depicted in Fig. 2. An individual wind turbine is typically rated for capacities 1-4 MW at around 690 V. A pad mounted generator step-up transformer usually steps up the generation voltage of 600-690 V to 34.5 kV by the transformer between buses 4 and 5. Multiple wind turbine models are connected at the 34.5 kV collector bus between buses 3 and 4. The operating voltage at the collector bus is further stepped up at the interconnection to the transmission voltage level at 132 kV or 230 kV via a substation transformer between buses 2 and 3. The representation in Fig. 2 is considered adequate for positive sequence dynamic simulations [21]. Type 4 wind energy generator (Wt4g), turbine (Wt4t) and exciter models (Wt4e) are used to represent the wind energy penetration. The power system simulator used to carry out dynamic simulations is GE PSLF.

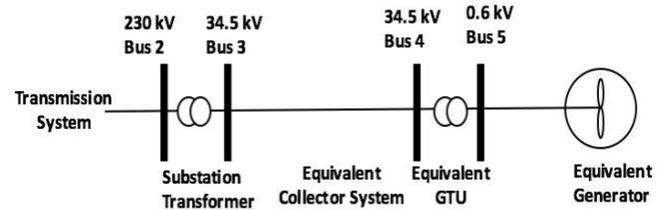

Fig. 2: Single line diagram of wind turbine [21].

### D. Supervised learning for islanding detection

The islanding detection algorithm in the proposed approach utilizes DTs. DT is a supervised learning-based data mining technique which infers hidden relationships from the data and classifies it based on binary partitioning through if-else statements [22]. In this paper, a Classification and Regression Tree (CART)-based DT is trained offline with the help of a training database and a mapping is developed by finding correlations between the input and the output. In [9], [22]-[24] it is observed that DT based classifiers detect island formation accurately and reliably. As such, DTs are used to evaluate the performance of the proposed methodology.

To create a robust dataset for accurate islanding detection, islanding and non-islanding scenarios were created and simulated in accordance with the following methodology.

1. *Generation of simulation cases*: For non-islanding scenarios, some extreme cases such as line trips, faults, and generator trips were simulated, and the measurement of voltage phase angle for these cases recorded from GE PSLF. For creating island in large test systems $i, 1 \leq i \leq 5$ transmission lines were removed at different instants of time using the community-based partitioning scheme developed in [25].
2. *Measurement of voltage phase angle*: For each case, the voltage phase angle measurements required for calculating CUSPAD values are obtained using the model *ametr* in GE PSLF. It is assumed that PMUs are installed on multiple locations in the system under study and the bus voltage angle measurements are provided by them; see Section II.E for the PMU placement methodology that was employed in this paper.

3. *Calculation of CUSPAD*: Dynamic simulations were run in GE PSLF to record the phase angle measurements at the rate of 30 samples per second to emulate PMU data reporting rate. CUSPAD is computed based on the methodology described in Section II.B.
4. *Training Data*: After the CUSPAD values for every simulation is obtained, they are fed as inputs to CART. Every case in the training dataset is identified as an islanding case or a non-islanding case by labeling it as 0 or 1 [22]. This serves as the training database for the DT.
5. *Testing Data*: To test the DT model built in the previous step, realistic measurements are replicated through introduction of measurement errors in the training database. The error model used is additive and includes both PMU and instrumentation channel errors:
   i. PMU errors in voltage phase angles are assumed to be a Gaussian distribution with zero mean and standard deviation of 0.104° [12].
   ii. Instrumentation channel errors in voltage phase angle are assumed to follow a uniform distribution that lies in the range of ±1°, ±2°, or ± 4° for the different case studies considered. Good quality measurements (for example, revenue quality instrument transformers) are also considered for testing purpose. They are assumed to introduce an angle error of the order of 0.1° [12].

The resultant voltage phase angles after incorporation of additive PMU and instrumentation errors is given by [26]:
$$\theta_V^m = \theta_V^t + \alpha_{VT}^{error} + \alpha_{PMU}^{error} \tag{8}$$
where $\theta_V^m$ is the measured voltage phase angle and $\theta_V^t$ is the true voltage phase angle. The instrumentation channel errors are denoted by $\alpha_{VT}^{error}$ while the PMU errors are denoted by $\alpha_{PMU}^{error}$. A schematic diagram describing the different steps that were followed for training and testing the DT-based islanding detection classifier is shown in Fig. 3.

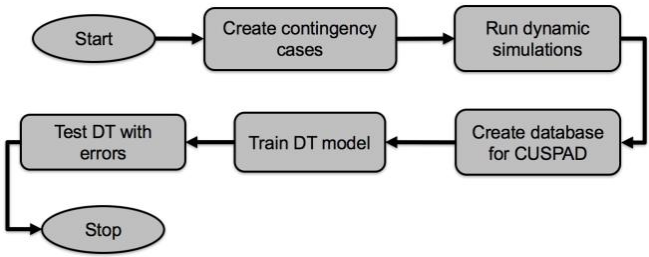

Fig. 3: Flowchart for the proposed CUSPAD approach.

### E. PMU placement

When PMUs are placed in a network, the primary objective is to ensure observability, i.e. the PMUs should have the ability to directly or indirectly observe all the bus voltages of the network. In addition to ensuring topological observability, the PMU placement scheme proposed in [27] takes into consideration PMU redundancy for critical buses as well as the cost of disrupting a substation for PMU installation. Accordingly, in this paper, the core concept of [27] is employed for determining the locations where PMUs must be placed.

Let the power network be denoted by an undirected graph $G(V, E)$ such that $V$ is the set of nodes (buses) and $E$ is the set of edges (transmission lines or transformers). The buses are grouped into substations, $S$, using the rationale that buses connected by transformers will lie inside the same substation. It is assumed in this study that all PMUs are of the dual-use line relay (DULR)-type. For each substation $S_i \in S$, a binary variable $x_i$ is used such that the following holds true:
$$x_i = \begin{cases} 1, & \text{if Substation } S_i \text{ is distrupted} \\ 0, & \text{otherwise} \end{cases} \tag{9}$$
Each edge $e \in E$ is associated with two binary variables $w_e^l$ and $w_e^h$ such that following holds true:
$$w_e^l = \begin{cases} 1, & \text{if DULR is placed at the low end of edge } e \\ 0, & \text{otherwise} \end{cases} \tag{10}$$
$$w_e^h = \begin{cases} 1, & \text{if DULR is placed at the high end of edge } e \\ 0, & \text{otherwise} \end{cases} \tag{11}$$
The objective is to minimize the total cost of PMU installations which involve cost of PMU devices as well as the cost of disrupting a substation. This objective function is mathematically described by:
$$Minimize \left( \sum_{i=1}^{k} c_i x_i + \Delta \sum_{e \in E} \{w_e^h + w_e^l\} \right) \tag{12}$$
where, $c_i$ is the cost of disrupting a substation, $\Delta$ is the cost of a DULR, and $k = |S|$. If $E_v$ denotes all outgoing phases from a vertex $v$, the constraint for phase observability is given by:
$$\sum_{e \in E_v} \{w_e^h + w_e^l\} \geq 1 \tag{13}$$

### III. SIMULATION AND RESULTS

In this section, the efficiency of CUSPAD in islanding detection is compared with that of the conventional AD approach. For the AD approach, pairs of voltage phase angle differences are calculated through instantaneous combinations of PMU measurements. The test systems comprised of a modified version of the 18-bus system available in the GE PSLF library and the IEEE 118-bus system. Measurement errors consisting of both PMU and instrumentation channel errors were included in the test data. The error model used for the two error types can be found in Section II.D. The simulations were repeated 50 times and a 95% confidence interval accuracy was computed for the test data.

#### A. Modified 18-bus test case

The original 18-bus system is modified to include wind energy penetration in the network by replacing one of the conventional generators with an equivalent capacity wind farm connected at the 230 kV voltage level. The number of PMUs required for complete observability of the 18-bus system was 5 and they were located at buses 1, 11, 14, 23, and 31. Total cases simulated were 467 out of which 200 were islanding cases and 267 were non-islanding cases. To determine a suitable window length for calculating CUSPAD, the DT accuracies obtained with various window sizes are presented in Fig. 4. We observe that the relative increase in DT accuracy corresponding to the window size between 30 and 40 samples is less as compared to that obtained for window sizes between 20 and 30 samples. A larger window size would however negatively influence the detection time (by adding more delay). As with any islanding detection algorithm, a lower detection time is preferred and therefore a compromise between DT accuracy and window size must be made. In the literature, a time delay of 100-150 ms was considered in [13] to prevent misclassifications. Islanding detection time as high as 2-3 seconds is discussed in [4]. Taking all this into account, we believe that for the proposed study, a window size of 30 samples would be appropriate. Comparing accuracies in Table I, it can be concluded that for the 18-bus system, for a window-size of 30 samples, the CUSPAD approach is not affected by increasing amounts of additive measurement errors while the performance of the conventional AD approach deteriorates considerably as the errors increase in the measurements.

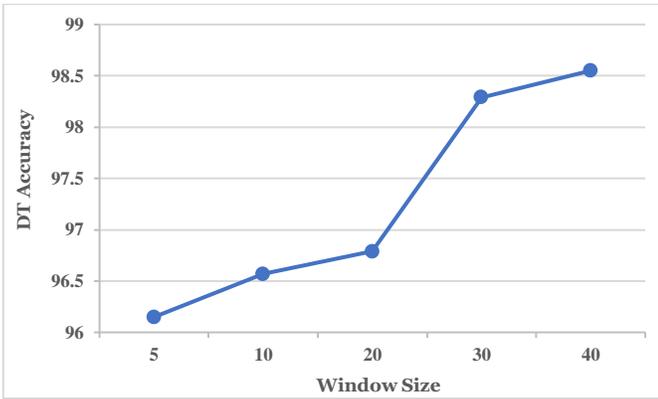

Fig. 4: Selection of window size for CUSPAD calculation

**Table I: Accuracy comparison of DT models for modified 18 bus system (16% wind penetration)**

| Error | | AD | | CUSPAD | |
|---|---|---|---|---|---|
| Instrumentation Channel Error | PMU Error | Accuracy (95%) | Depth | Accuracy (95%) | Depth |
| 0° | 0 Mean ±0.10 4° SD | 99.79 | 5 | 98.29 | 5 |
| $-0.1° \leq \alpha_{VT}^{error} \leq 0.1°$ | | 99.37 ± 0.06 | 5 | 98.01 ± 0.07 | 5 |
| $-1° \leq \alpha_{VT}^{error} \leq 1°$ | | 96.85 ± 0.51 | 5 | 98.06 ± 0.07 | 5 |
| $-2° \leq \alpha_{VT}^{error} \leq 2°$ | | 94.66 ± 0.68 | 5 | 98.08 ± 0.06 | 5 |
| $-4° \leq \alpha_{VT}^{error} \leq 4°$ | | 93.04 ± 380 | 5 | 97.97 ± 0.08 | 5 |

### B. Modified 118-bus test case

The original IEEE 118-bus system is modified to include variable percentages of wind energy penetration in the network by replacing some of the conventional generators with equivalent capacity wind farms connected at the 132 kV voltage level. The number of PMUs required for complete observability of this system was 38. They were placed on buses 3, 5, 8, 9, 12, 15, 17, 21, 23, 28, 30, 36, 40, 43, 45, 49, 52, 56, 59, 63, 65, 66, 68, 69, 71, 75, 77, 80, 85, 86, 84, 91, 94, 101, 105, 110, 114, and 116. To create islands in the IEEE 118-bus system, the community-based partitioning logic developed in [25] was used. It identifies the minimum number of edges that must be lost for islands of a given size to form. In total, 2,000 cases were simulated for three levels of wind penetration, namely, 10%, 20%, and 30%. Of these 2,000 cases, 1,000 were islanding cases and 1,000 were non-islanding cases. The 30-sample window size was also selected for computing accuracy of CUSPAD for the 118-bus system. The results obtained for 30% wind penetration are presented in Table II. A comparison between AD and CUSPAD accuracies for 10%-30% wind penetration is depicted as a plot in Fig. 5. It is evident from Table II and Fig. 5 that the performance of CUSPAD is superior to AD in presence of increasing amounts of additive instrumentation channel errors.

**Table II: Accuracy comparison of DT models for modified 118 bus system (30% wind penetration)**

| Error | | AD | | CUSPAD | |
|---|---|---|---|---|---|
| Instrumentation Channel Error | PMU Error | Accuracy (95%) | Depth | Accuracy (95%) | Depth |
| 0° | 0 Mean ±0.10 4° SD | 99.80 | 4 | 99.80 | 4 |
| $-0.1° \leq \alpha_{VT}^{error} \leq 0.1°$ | | 93.45 ± 2.79 | 4 | 99.20 ± 0.14 | 4 |
| $-1° \leq \alpha_{VT}^{error} \leq 1°$ | | 78.40 ± 2.12 | 4 | 99.24 ± 0.15 | 4 |
| $-2° \leq \alpha_{VT}^{error} \leq 2°$ | | 69.06 ± 2.40 | 4 | 99.22 ± 0.15 | 4 |
| $-4° \leq \alpha_{VT}^{error} \leq 4°$ | | 66.18 ± 3.12 | 4 | 99.22 ± 0.15 | 4 |

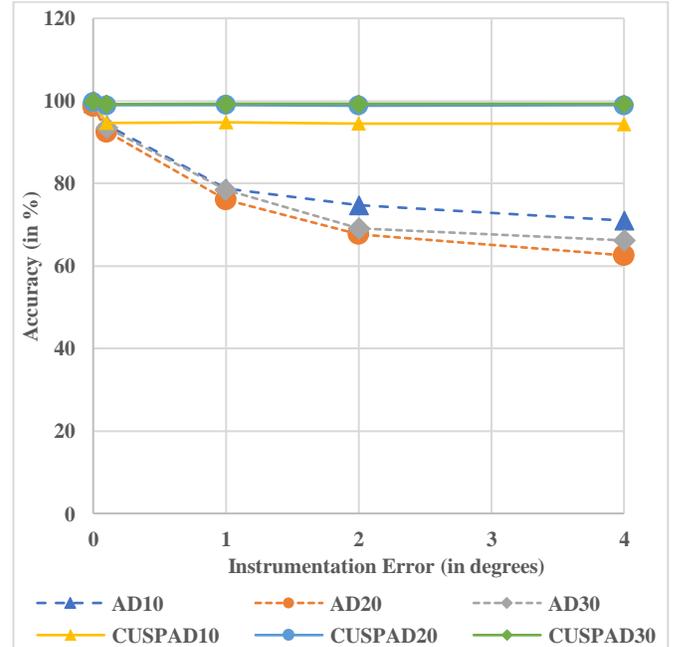

Fig. 5: Pictorial Analysis of AD and CUSPAD accuracies for 10% - 30% wind energy penetration

## IV. Conclusions

In this paper, a PMU based passive islanding detection technique that is immune to additive instrumentation channel errors is proposed and tested in the presence of different levels of wind penetration. The cumulated sum of voltage phase angle difference (CUSPAD) obtained from a specific PMU device over a given time-period cancels the effect of instrumentation channel errors present in the PMU measurements. This is the underlying reason behind consistent islanding detection accuracy in presence of increasing instrumentation channel errors. The proposed approach has been tested for an 18-bus test system using DT based CART classifier. The results indicate that, in presence of instrumentation channel errors, the proposed CUSPAD technique is superior to the conventional AD approach. The performance of this technique is further evaluated for the IEEE 118-bus system where 10%, 20%, and 30% wind penetration is modeled by replacing corresponding synchronous generation. The performance of the CUSPAD approach was also found to be superior for the 118-bus system in presence of increasing amounts of instrumentation channel errors when compared to that of the AD approach. We can therefore conclude that islanding detection in renewable rich systems using CUSPAD is more reliable than AD in presence of additive instrumentation channel errors.

## Acknowledgements

This work is partially funded by the Power Systems Engineering Research Center (PSERC) project, S-74. The authors would also like to acknowledge the valuable and constructive feedback provided by Alan Engelmann of ComEd during this research.